\documentclass{llncs}

\usepackage{amsmath}
\usepackage{amssymb}
\usepackage[dvips]{color}
\usepackage[dvips]{graphicx}
\usepackage{enumerate}

\begin{document}

\pagestyle{plain} \mainmatter

\title{Strongly Multiplicative and 3-Multiplicative Linear Secret Sharing Schemes 
}



\author{Zhifang Zhang\inst{1} \and Mulan Liu\inst{1}\and Yeow Meng Chee
\inst{2}\and San Ling\inst{2}\and Huaxiong Wang\inst{2,3}}


\institute{Key Laboratory of Mathematics Mechanization, Academy of
Mathematics and Systems Science, Chinese Academy of Sciences,
Beijing, China \\\email{\{zfz, mlliu\}@amss.ac.cn} \and Division of
Mathematical Sciences, School of Physical and Mathematical Sciences,
Nanyang Technological University, Singapore
\\\email{\{ymchee, lingsan, hxwang\}@ntu.edu.sg} 
\and
Centre for Advanced Computing - Algorithms and Cryptography\\
Department of Computing\\
Macquarie University, Australia}


\maketitle

\begin{abstract}
Strongly multiplicative linear secret sharing schemes (LSSS) have
been a powerful tool for constructing secure multi-party
computation protocols. However, it remains open {\it whether or not there
exist  efficient constructions of strongly multiplicative LSSS from
general LSSS}. In this paper, we propose the new concept of a
{\it $3$-multiplicative LSSS}, and establish its relationship with
strongly multiplicative LSSS. More precisely, we show that any
3-multiplicative LSSS is a strongly multiplicative LSSS, but the
converse is not true; and that any strongly multiplicative LSSS can
be efficiently converted into a 3-multiplicative LSSS. Furthermore,
we apply 3-multiplicative LSSS to the computation of unbounded
fan-in multiplication, which reduces its round complexity to four (from
five of the previous protocol based on strongly multiplicative LSSS).
We also give two constructions of 3-multiplicative LSSS from
Reed-Muller codes and algebraic geometric codes. We believe that the
construction and verification of 3-multiplicative LSSS are easier
than those of strongly multiplicative LSSS. This presents a step forward in
settling the open problem of efficient constructions of strongly multiplicative LSSS from
general LSSS.

\end{abstract}

{\small{\bf Keywords~~ }monotone span
program, secure multi-party computation, strongly multiplicative linear secret
sharing scheme}

\section{Introduction}
Secure multi-party computation (MPC)
\cite{Yao:Protocols,Goldreich:How} is a cryptographic primitive that enables
$n$ players to jointly compute an agreed function of their private
inputs in a secure way, guaranteeing the correctness of the outputs
as well as the privacy of the players' inputs, even when some
players are malicious. It has become a fundamental tool in
cryptography and distributed computation.  Linear secret sharing
schemes (LSSS) play an important role in building MPC
protocols. Cramer {\em et al.} \cite{Cramer:General}
developed a generic method of constructing MPC protocols from LSSS.
Assuming that the function to be computed is represented as an
arithmetic circuit over a finite field, their protocol ensures that each
player share his private input through an LSSS, and then evaluates
the circuit gate by gate. The main idea of their protocol is to keep
the intermediate results secretly shared among the players with the
underlying  LSSS. Due to the nature of linearity, secure additions
(and linear operations) can be easily achieved. For instance, if
player $P_i$ holds the share $x_{1i}$ for input $x_1$ and
$x_{2i}$ for input $x_2$, he can locally compute $x_{1i}+x_{2i}$ which is
actually $P_i$'s share for $x_1+x_2$. Unfortunately, the above
homomorphic property does not hold for multiplication. In order to
securely compute multiplications, Cramer {\em et al.}
\cite{Cramer:General} introduced the concept of {\it multiplicative} LSSS, where
the product $x_1x_2$ can be computed as a linear combination of the
local products of shares, that is, $x_1x_2=\sum_{i=1}^na_ix_{1i}x_{2i}$
for some constants $a_i, 1\leq i\leq n$. Since $x_{1i}x_{2i}$ can be
locally computed by $P_i$, the product can then be securely
computed through a linear combination. Furthermore, in order to
resist against an active adversary, they defined {\it strongly}
multiplicative LSSS,  where $x_1x_2$ can be computed as a linear
combination of the local products of shares by all players excluding
any corrupted subset. Therefore,  multiplicativity becomes an
important property in constructing secure MPC protocols. For
example, using strongly multiplicative LSSS, we can construct an
error-free MPC protocol secure against an active adversary in the
information-theoretic model \cite{Cramer:General}. Cramer {\em et al.}
\cite{Cramer:OnCodes} also gave an efficient
reconstruction algorithm for strongly multiplicative LSSS that
recovers the secret even when the shares submitted by the corrupted
players contain errors. This implicit ``built-in'' verifiability
makes strongly multiplicative LSSS an attractive building block
for MPC protocols.

Due to their important role as the building blocks in MPC protocols,
efficient constructions of multiplicative LSSS and strongly
multiplicative LSSS have been studied by several authors in recent
years. Cramer {\em et al.} \cite{Cramer:General} developed a generic
method of constructing a multiplicative LSSS from any given LSSS
with a double expansion of the shares. Nikov {\em et al.}
\cite{Nikov:multiplicativeLSSS} studied how to securely compute
multiplications in a dual LSSS,  without blowing up the shares.
For some specific access structures there exist very efficient
multiplicative LSSS.  Shamir's threshold secret sharing scheme is a
well-known example of an ideal (strongly) multiplicative LSSS.
Besides, self-dual codes give rise to ideal multiplicative
LSSS \cite{Cramer:OnCodes}, and Liu {\em et al.} \cite{Liu:IEEE} provided
a further class of ideal multiplicative LSSS for graph access
structures. We note that for strongly multiplicative LSSS, the
known general construction is of exponential complexity. K\"{a}sper
{\em et al.} \cite{KNN:SMHTSS} gave some efficient constructions for
specific access structures (hierarchical threshold structures). It
remains open whether there exists an efficient transformation from a
general LSSS to a strongly multiplicative one.

On the other hand, although in a multiplicative LSSS, multiplication
can be converted into a linear combination of inputs from the
players, each player has to {\em reshare} the product of his shares, that is,
for $1\leq i\leq n$, $P_i$ needs to reshare the product
$x_{1i}x_{2i}$ to securely compute the linear combination
$\sum_{i=1}^na_ix_{1i}x_{2i}$. This resharing process involves
costly interactions among the players. For example,  if the players
are to securely compute multiple multiplications,
$\prod_{i=1}^lx_i$, the simple sequential multiplication requires
interaction of round complexity proportional to $l$.  Using the
technique developed by Bar-Ilan and Beaver \cite{BB:ConstantRd}, Cramer {\em et al.} \cite{CKP:LinearAlgbra} recently showed that the round complexity can be
significantly reduced to a constant of five for unbounded fan-in
multiplications. However, the method does not seem efficient when
$l$ is small. For example, considering  $x_1x_2$ and $x_1x_2x_3$, extra
rounds of interactions seem unavoidable for computing $x_1x_2x_3$ if
we apply the method of Cramer {\em et al.} \cite{CKP:LinearAlgbra}.

\subsection{Our Contribution}

In this paper, we propose the concept of 3-multiplicative LSSS.
Roughly speaking, a 3-multiplicative LSSS is a generalization of
multiplicative LSSS, where the product $x_1x_2x_3$ is a linear
combination of the local products of shares. As one would expect, a
3-multiplicative LSSS achieves better round complexity for the computation of
$\prod_{i=1}^lx_i$ compared to a multiplicative LSSS, if
$l \geq 3$. Indeed, it is easy to see that computing the
product $\prod_{i=1}^9x_i$ requires two rounds of interaction for a
3-multiplicative LSSS but four rounds for a multiplicative LSSS. We
also extend the concept of a 3-multiplicative LSSS to the more general
$\lambda$-multiplicative LSSS, for all integers $\lambda \geq 3$, and
show that $\lambda$-multiplicative LSSS reduce the round
complexity by a factor of $\frac{1}{\log\lambda}$ from multiplicative
LSSS. In particular, 3-multiplicative LSSS reduce the constant
round complexity of computing the unbounded fan-in multiplication
from five to four, thus improving a result of Cramer {\em et al.} \cite{CKP:LinearAlgbra}.

More importantly, we show that 3-multiplicative LSSS are
closely related to strongly multiplicative LSSS. The latter is
known to be a powerful tool for constructing secure MPC protocols
against active adversaries. More precisely, we show the following: 
\begin{enumerate}[(i)]
\item 3-multiplicative LSSS are also strongly multiplicative;
\item there exists an efficient algorithm that transforms a strongly
multiplicative LSSS into a 3-multiplicative LSSS; 
\item an example of a strongly multiplicative LSSS that is not 3-multiplicative.
\end{enumerate}

Our results contribute to the study of MPC in the following three
aspects:
\begin{itemize}
\item The 3-multiplicative LSSS outperform strongly
multiplicative LSSS with respect to round complexity in
the construction of secure MPC protocols.

\item The 3-multiplicative LSSS are easier to construct than
strongly multiplicative LSSS. First, the existence of an efficient transformation
from a strongly multiplicative LSSS to a 3-multiplicative LSSS
implies that efficiently constructing 3-multiplicative LSSS is
not a harder problem. Second, verification of a strongly
multiplicative LSSS requires checking the linear combinations for
all possibilities of adversary sets, while the verification of a
3-multiplicative LSSS requires only one checking. We give two
constructions of LSSS based on Reed-Muller codes and algebraic geometric
codes that can be easily verified for 3-multiplicativity,
 but it does not seem easy to give direct proofs of their strong multiplicativity.

\item  This work provides two possible directions toward solving
the open problem of determining the
existence of efficient constructions for strongly multiplicative
LSSS. On the negative side, if we can prove that in the
information-theoretic model and with polynomial size message
exchanged, computing $x_1x_2x_3$ inevitably needs more rounds of
interactions than computing $x_1x_2$, then we can give a negative
answer to this open problem. On the positive side, if we can find an
efficient construction for 3-multiplicative LSSS, which also
results in strongly multiplicative LSSS, then we will have an
affirmative answer to this open problem.

\end{itemize}

\subsection{Organization}

Section \ref{sec:pre} gives notations, definition of
multiplicative LSSS, and general constructions for strongly
multiplicative LSSS. 
Section \ref{sec:3-m} defines
3-multiplicative LSSS. 
Section \ref{sec:3-strong} shows the
relationship between 3-multiplicative LSSS and strongly multiplicative
LSSS. Section \ref{sec:constructions} gives two constructions of 3-multiplicative
LSSS from
error-correcting codes, and Section \ref{sec:implication} discusses
the implications of 3-multiplicative LSSS in MPC. Section
\ref{sec:conclusion} concludes the paper.

\section{Preliminaries} \label{sec:pre}

Throughout this paper, let $P=\{P_1,\ldots,P_n\}$ denote the set of $n$
players and let $\mathcal{K}$ be a finite field. In a secret sharing
scheme, the collection of all subsets of players that are authorized to recover
the secret is called its {\it access structure}, and is denoted $AS$.
An access structure possesses the monotone ascending property: if $A'\in AS$, then
for all $A\subseteq P$ with $A\supseteq A'$, we also have $A\in AS$.
Similarly, the collection of subsets of players that are possibly corrupted
is called the {\it adversary structure}, and is denoted $\mathcal{A}$.
An adversary structure possesses the monotone descending property: if $A'\in
\mathcal{A}$, then for all $A\subseteq P$ with $A\subseteq A'$, we also have
$A\in \mathcal{A}$. Owing to these monotone properties, it is
often sufficient to consider the {\it minimum access
structure} $AS_{min}$ and the {\it maximum adversary structure}
$\mathcal{A}_{max}$ defined as follows:
\begin{align*}
AS_{min} &=\{A\in AS\mid\forall B\subseteq P, \text{ we have } B\subsetneq A\Rightarrow B\not\in
AS\},\\
\mathcal{A}_{max} &=\{A\in \mathcal{A}\mid
\forall B\subseteq P, \text{ we have } B\supsetneq A\Rightarrow B\not\in\mathcal{A}\}.
\end{align*}
In this paper, we consider the {\em complete} situation, that is,
$\mathcal{A}=2^P-AS$. Moreover, an adversary structure
$\mathcal{A}$ is called $Q^2$ (respectively, $Q^3$) if any two (respectively,
three) sets in $\mathcal{A}$ cannot cover the entire player set
$P$. For simplicity, when an adversary structure $\mathcal{A}$ is
$Q^2$ (respectively, $Q^3$) we also say the corresponding access structure
$AS=2^P-\mathcal{A}$ is $Q^2$ (respectively, $Q^3$).

\subsection{Linear Secret Sharing Schemes and Monotone Span Programs}

Suppose $S$ is the secret-domain, $R$ is the set of random inputs,
and $S_i$ is the share-domain of $P_i$, where $1\leq i\leq n$. 
Let {\sf S} and {\sf R} denote random variables taking values in $S$ and $R$, respectively.
Then
$\Pi:S\times R\rightarrow S_1\times\cdots\times S_n$ is called a {\it
secret sharing scheme} (SSS)  with respect to the access structure $AS$, if
the following two conditions are satisfied:
\begin{enumerate}
\item for all $A\in AS$,  $H({\sf S}\mid \Pi({\sf S},{\sf R})|_A)=0$;
\item  for all $B\not\in AS$, $H({\sf S}\mid \Pi({\sf S},{\sf R})|_B)=H({\sf S})$,
\end{enumerate}
where $H(\cdot)$ is the entropy function. Furthermore, the secret
sharing scheme $\Pi$ is called {\it linear} if we have
$S=\mathcal{K}$, $R=\mathcal{K}^{l-1}$, and $S_i=\mathcal{K}^{d_i}$ for
some positive integers $l$ and $d_i$, $1\leq i\leq n$, and the
reconstruction of the secret can be performed by taking a linear
combination of shares from the authorized players. The quantity $d =
\sum_{i=1}^nd_i$ is called the {\it size} of the LSSS.

Karchmer and Wigderson \cite{Karchmer:On} introduced monotone
span programs (MSP) as a linear model for computing
monotone Boolean functions. We denote an MSP by
$\mathcal{M}(\mathcal{K},M,\psi,\vec{v})$, where $M$ is a $d\times
l$ matrix over $\mathcal{K}$,
$\psi:\{1,\ldots,d\}\rightarrow\{P_1,\ldots,P_n\}$ is a surjective
labeling map, and $\vec{v}\in\mathcal{K}^l$ is a nonzero vector. We
call $d$ the {\it size} of the MSP and $\vec{v}$ the {\it target
vector}. A monotone Boolean function $f:\{0,1\}^n\rightarrow
\{0,1\}$ satisfies $f(\vec{\delta}')\geq f(\vec{\delta})$ for
any $\vec{\delta}'\geq\vec{\delta}$, where
$\vec{\delta}=(\delta_1,\ldots,\delta_n)$,
$\vec{\delta}'=(\delta'_1,\ldots,\delta'_n)\in \{0,1\}^n$, and
$\vec{\delta}'\geq\vec{\delta}$ means $\delta'_i\geq\delta_i$ for
$1\leq i\leq n$. We say that an MSP
$\mathcal{M}(\mathcal{K},M,\psi,\vec{v})$ {\it computes the monotone
Boolean function $f$} if $\vec{v}\in span\{M_A\}$ if
and only if $f(\vec{\delta_A})=1$, where $A$ is a set of players, $M_A$ denotes the matrix
constricted to the rows labeled by players in $A$, $span\{M_A\}$
denotes the linear space spanned by the row vectors of $M_A$, and
$\vec{\delta_A}$ is the characteristic vector of $A$.

\begin{theorem}[Beimel \cite{Beimel:Secure}] \label{thLSSS&MSP}
Suppose $AS$ is an access structure over $P$ and $f_{AS}$ is the
characteristic function of $AS$, that is, $f_{AS}(\vec{\delta})=1$ if
and only if $\vec{\delta}=\vec{\delta}_A$ for some $A\in AS$. Then
there exists an LSSS of size $d$ that realizes $AS$ if and only if
there exists an MSP of size $d$ that computes $f_{AS}$.
\end{theorem}

Since an MSP computes the same Boolean function under linear
transformations, we can always assume that the target vector is
$\vec{e}_1=(1,0,\ldots,0)$. From an MSP
$\mathcal{M}(\mathcal{K},M,\psi,\vec{e}_1)$ that computes $f_{AS}$,
we can derive an LSSS realizing $AS$ as follows: to share a secret
$s\in\mathcal{K}$, the dealer randomly selects
$\vec{\rho}\in\mathcal{K}^{l-1}$, computes $M(s,\vec{\rho})^\tau$
and sends $M_{P_i}(s,\vec{\rho})^\tau$ to $P_i$ as his share, where
$1\leq i\leq n$ and $\tau$ denotes the transpose.
The following property of MSP is useful in the proofs of our results.

\begin{proposition}[Karchmer and Wigderson \cite{Karchmer:On}]\label{prop1}
Let $\mathcal{M}(\mathcal{K},M,\psi,\vec{e_1})$ be an MSP that
computes a monotone Boolean function $f$. Then for all
$A\subseteq P$, $\vec{e}_1\not\in span\{M_A\}$ if and only if
there exists $\vec{\rho}\in\mathcal{K}^{l-1}$ such that
$M_A(1,\vec{\rho})^\tau=\vec{0}^\tau$.
\end{proposition}

\subsection{Multiplicative Linear Secret Sharing Schemes}

From Theorem \ref{thLSSS&MSP},  an LSSS can be identified with its
corresponding MSP in the following way. Let
$\mathcal{M}(\mathcal{K},M,\psi,\vec{e}_1)$ be an LSSS realizing
the access structure $AS$. Given two vectors
$\vec{x}=(x_1,\ldots,x_d)$,
$\vec{y}=(y_1,\ldots,y_d)\in\mathcal{K}^d$, we define
$\vec{x}\diamond \vec{y}$ to be the vector containing all entries
of the form $x_i\cdot y_j$ with $\psi(i)=\psi(j)$. More precisely,
let
\begin{align*}
\vec{x}&=(x_{11},\ldots,x_{1d_1},\ldots,x_{n1},\ldots,x_{nd_n}),\\
\vec{y}&=(y_{11},\ldots,y_{1d_1},\ldots,y_{n1},\ldots,y_{nd_n}),
\end{align*}
 where
$\sum_{i=1}^nd_i=d$,  and $(x_{i1},\ldots,x_{id_i})$,
$(y_{i1},\ldots,y_{id_i})$ are the entries distributed to $P_i$
according to $\psi$. Then $\vec{x}\diamond \vec{y}$ is the vector
composed of the $\sum_{i=1}^nd_i^{\;2}$ entries $x_{ij}y_{ik}$,
where $1\leq j,k\leq d_i, 1\leq i\leq n$.  For consistency, we
write the entries of $\vec{x}\diamond \vec{y}$ in some fixed
order.  We also define $(\vec{x}\diamond
\vec{y})^\tau=\vec{x}^\tau\diamond\vec{y}^\tau$.

\begin{definition}[Multiplicativity]\label{defMLSS}
Let $\mathcal{M}(\mathcal{K},M,\psi,\vec{e_1})$ be an LSSS realizing
the access structure $AS$ over $P$. Then $\mathcal{M}$ is called
{\em multiplicative} if there exists a recombination vector
$\vec{z}\in\mathcal{K}^{\sum_{i=1}^nd_i^2}$, such that for all
$s,s'\in\mathcal{K}$ and
$\vec{\rho},\vec{\rho}'\in\mathcal{K}^{l-1}$, we have
\begin{equation*}
ss'=\vec{z}(M(s,\vec{\rho})^\tau\diamond M(s',\vec{\rho}')^\tau).
\end{equation*}
Moreover, $\mathcal{M}$ is {\em strongly multiplicative} if for all
$A\in \mathcal{A}=2^P-AS$, $\mathcal{M}_{\overline{A}}$ is
multiplicative, where $\mathcal{M}_{\overline{A}}$ denotes the MSP
$\mathcal{M}$ constricted to the subset $\overline{A}=P-A$.
\end{definition}

\begin{proposition}[Cramer {\em et al.} \cite{Cramer:General}] \label{propQ3}
Let $AS$ be an access structure over $P$. Then there exists a
multiplicative (respectively, strongly multiplicative) LSSS realizing $AS$
if and only if $AS$ is $Q^2$ (respectively, $Q^3$).
\end{proposition}

\subsection{General Constructions of Strongly Multiplicative LSSS} \label{subsec:2.3}

For all $Q^2$ access structure $AS$, Cramer {\em et al.}
\cite{Cramer:General} gave an efficient construction to build a
multiplicative LSSS from a general LSSS realizing the same $AS$. It
remains {\it open} if we can {\it efficiently} construct a strongly
multiplicative LSSS from an  LSSS. However, there are general
constructions with exponential complexity, as described below.

Since Shamir's threshold secret sharing scheme is strongly
multiplicative for all $Q^3$ threshold access structure, a proper
composition of Shamir's threshold secret sharing schemes results in a general construction
for strongly multiplicative LSSS \cite{Cramer:General}. Here,
we give another general construction based on multiplicative
LSSS.

Let $AS$ be any $Q^3$ access structure and
$\mathcal{M}(\mathcal{K},M,\psi,\vec{e}_1)$ be an LSSS realizing
$AS$. For all $A\in \mathcal{A}=2^P-AS$, it is easy to see that
$\mathcal{M}_{\overline{A}}$ realizes the restricted  access
structure $AS_{\overline{A}}=\{B\subseteq \overline{A}\mid B\in
AS\}$. The access structure $AS_{\overline{A}}$ is $Q^2$ over
$\overline{A}$ because $AS$ is $Q^3$ over $\overline{A}\cup A$.
Thus, we can transform $\mathcal{M}_{\overline{A}}$ into a
multiplicative LSSS following the general construction of Cramer {\em et al.}
\cite{Cramer:General} to obtain a strongly
multiplicative LSSS realizing $AS$. The example in Section \ref{subsec:example}
gives an illustration of this method.

We note that both constructions above give LSSS of exponential
sizes,
and hence are not {\em efficient} in general.

\section{3-Multiplicative and $\lambda$-Multiplicative LSSS} \label{sec:3-m}

In this section, we give an equivalent definition for
(strongly) multiplicative LSSS. We then define 3-multiplicative
LSSS and give a necessary and sufficient condition for its
existence. The notion of 3-multiplicativity is also extended to
$\lambda$-multiplicativity for all integer $\lambda>1$. Finally, we present a
generic (but inefficient) construction of $\lambda$-multiplicative
LSSS.

Under the same notations used in Section 2.2,  it is straightforward
to see that we have an induced labeling map
$\psi':\{1,\ldots,\sum_{i=1}^nd_i^2\}\rightarrow\{P_1,\ldots,P_n\}$ on the
entries of $\vec{x}\diamond\vec{y}$, distributing the entry
$x_{ij}y_{ik}$ to $P_i$, since both $x_{ij}$ and $y_{ik}$ are labeled
by $P_i$ under $\psi$. For an MSP
$\mathcal{M}(\mathcal{K},M,\psi,\vec{e}_1)$, denote
$M=(M_1,\ldots,M_l)$, where $M_i\in\mathcal{K}^d$ is the $i$-th column
vector of $M$, $1\leq i\leq l$. We construct a new matrix
$M_{\diamond}$ as follows:
\begin{equation*}
M_{\diamond}=(M_1\diamond M_1,\ldots,M_1\diamond M_l,M_2\diamond M_1,\ldots,M_2\diamond M_l,\ldots,M_l\diamond M_1,\ldots,M_l\diamond M_l).
\end{equation*}
For consistency, we also denote $M_{\diamond}$ as $M\diamond M$.
Obviously, $M_{\diamond}$ is a matrix over $\mathcal{K}$ with
$\sum_{i=1}^nd_i^2$ rows and $l^2$ columns. For any two vectors
$\vec{u},\vec{v}\in \mathcal{K}^l$, it is easy to verify that
\begin{equation*}
(M\vec{u}^\tau)\diamond(M\vec{v}^\tau)=M_{\diamond}(\vec{u}\otimes\vec{v})^\tau,
\end{equation*}
where $\vec{u}\otimes\vec{v}$ denotes the tensor product with its
entries written in a proper order. Define the induced labeling map
$\psi'$ on the rows of $M_{\diamond}$. We have the following
proposition.

\begin{proposition}\label{prop2}
Let $\mathcal{M}(\mathcal{K},M,\psi,\vec{e}_1)$ be an LSSS
realizing the access structure $AS$, and let $M_{\diamond}$ be with
the labeling map $\psi'$. Then $\mathcal{M}$ is
multiplicative if and only if $\vec{e}_1\in span\{M_{\diamond}\}$,
where $\vec{e}_1 = (1, 0, \dots, 0)$. Moreover, $\mathcal{M}$ is
strongly multiplicative if and only if $\vec{e}_1\in
span\{(M_{\diamond})_{\overline{A}}\}$ for all
$A\in\mathcal{A}=2^P-AS$.
\end{proposition}

\begin{proof}
By Definition \ref{defMLSS}, $\mathcal{M}$ is multiplicative if and
only if $ss'=\vec{z}(M(s,\vec{\rho})^\tau\diamond
M(s',\vec{\rho}')^\tau)$ for all $s,s'\in\mathcal{K}$ and
$\vec{\rho},\vec{\rho}'\in\mathcal{K}^{l-1}$. Obviously,
\begin{equation}\label{eq-product}
M(s,\vec{\rho})^\tau\diamond
M(s',\vec{\rho}')^\tau=M_{\diamond}((s,\vec{\rho})\otimes(s',\vec{\rho}'))^\tau=M_{\diamond}(ss',\vec{\rho}'')^\tau,
\end{equation}
where $(ss',\vec{\rho}'')=(s,\vec{\rho})\otimes(s',\vec{\rho}')$. On
the other hand, $ss'=\vec{e}_1(ss',\vec{\rho}'')^\tau$. Thus
$\mathcal{M}$ is multiplicative if and only if
\begin{equation}\label{eq11}
(\vec{e}_1-\vec{z}M_{\diamond})(ss',\vec{\rho}'')^\tau=0.
\end{equation}
Because of the arbitrariness of $s,s',\vec{\rho}$ and $\vec{\rho}'$,
equality (\ref{eq11}) holds if and only if
$\vec{e}_1-\vec{z}M_{\diamond}=\vec{0}$. Thus $\vec{e}_1\in
span\{M_{\diamond}\}$. The latter part of the proposition can be
proved similarly.
$\Box$
\end{proof}

Now we are ready to give the definition of 3-multiplicative LSSS.
We extend the diamond product ``$\diamond$" and define
$\vec{x}\diamond\vec{y}\diamond\vec{z}$ to be the vector
containing all entries of the form $x_i y_jz_k$ with
$\psi(i)=\psi(j)=\psi(k)$, where the entries of
$\vec{x}\diamond\vec{y}\diamond\vec{z}$ are written in some fixed
order.

\begin{definition}[3-Multiplicativity]\label{def3M}
Let $\mathcal{M}(\mathcal{K},M,\psi,\vec{e}_1)$ be an LSSS realizing
the access structure $AS$. Then $\mathcal{M}$ is called
{\em $3$-multiplicative} if there exists a recombination vector
$\vec{z}\in\mathcal{K}^{\sum_{i=1}^nd_i^3}$ such that for all
$s_1,s_2,s_3\in\mathcal{K}$ and
$\vec{\rho}_1,\vec{\rho}_2,\vec{\rho}_3\in\mathcal{K}^{l-1}$, we have
\begin{equation*}
s_1s_2s_3=\vec{z}(M(s_1,\vec{\rho}_1)^\tau\diamond M(s_2,\vec{\rho}_2)^\tau\diamond M(s_3,\vec{\rho}_3)^\tau).
\end{equation*}
\end{definition}

We can derive an equivalent
definition for 3-multiplicative LSSS, similar to Proposition \ref{prop2}: $\mathcal{M}$ is
3-multiplicative if and only if $\vec{e}_1\in span\{(M\diamond
M\diamond M)\}$. The following proposition gives a necessary and
sufficient condition for the existence of 3-multiplicative
LSSS.

\begin{proposition}
For all access structures  $AS$,  there exists a
$3$-multiplicative LSSS realizing $AS$ if and only if $AS$ is $Q^3$.
\end{proposition}
\begin{proof}
Suppose $\mathcal{M}(\mathcal{K},M,\psi,\vec{e}_1)$ is a
3-multiplicative LSSS realizing $AS$, and suppose to the contrary, that
$AS$ is not $Q^3$, so there exist
$A_1,A_2,A_3\in\mathcal{A}=2^P-AS$ such that $A_1\cup A_2\cup
A_3=P$. By Proposition \ref{prop1}, there exists
$\vec{\rho}_i\in\mathcal{K}^{l-1}$ such that
$M_{A_i}(1,\vec{\rho}_i)^\tau=\vec{0}^\tau$ for $1\leq i\leq 3$.
Since $A_1\cup A_2\cup A_3=P$, we have $M(1,\vec{\rho}_1)^\tau\diamond
M(1,\vec{\rho}_2)^\tau\diamond M(1,\vec{\rho}_3)^\tau=\vec{0}^\tau$,
which contradicts Definition \ref{def3M}.

On the other hand, a general construction for building a
3-multiplicative LSSS from a strongly multiplicative LSSS is given
in the next section, thus sufficiency is guaranteed by
Proposition \ref{propQ3}.
$\Box$
\end{proof}

A trivial example of 3-multiplicative LSSS is Shamir's
threshold secret sharing scheme that realizes any $Q^3$ threshold
access structure. Using an identical argument for the case of
strongly multiplicative LSSS,  we have a general
construction for 3-multiplicative LSSS based on Shamir's threshold secret sharing
schemes, with exponential complexity.

For any $\lambda$ vectors
$\vec{x}_i=(x_{i1},\ldots,x_{id})\in\mathcal{K}^d,1\leq i\leq \lambda$,
we define $\diamond_{i=1}^{\lambda}\vec{x}_i$ to be the
$\sum_{i=1}^n{d_i^{\lambda}}$-dimensional vector which contains
entries of the form $\prod_{i=1}^{\lambda}x_{ij_i}$ with
$\psi(j_1)=\cdots=\psi(j_{\lambda})$.

\begin{definition}[\boldmath $\lambda$-Multiplicativity]
Let $\mathcal{M}(\mathcal{K},M,\psi,\vec{e}_1)$ be an LSSS
realizing the access structure $AS$, and let $\lambda>1$ be an integer.
Then $\mathcal{M}$ is {\em $\lambda$-multiplicative} if there
exists a recombination vector $\vec{z}$ such that for all
$s_1,\ldots,s_{\lambda}\in\mathcal{K}$ and
$\vec{\rho}_1,\ldots,\vec{\rho}_{\lambda}\in\mathcal{K}^{l-1}$, we
have
\begin{equation*}
\prod_{i=1}^{\lambda}s_i=\vec{z}(\diamond_{i=1}^{\lambda}M(s_i,\vec{\rho}_i)^\tau).
\end{equation*}
Moreover, $\mathcal{M}$ is {\em strongly
$\lambda$-multiplicative} if for all $A\not\in AS$, the
constricted LSSS $\mathcal{M}_{\overline{A}}$ is
$\lambda$-multiplicative.
\end{definition}

Again, we can define a new matrix by taking the diamond product of
$\lambda$ copies of $M$. This gives an equivalence to
(strongly) $\lambda$-multiplicative LSSS. Also, since Shamir's
threshold secret sharing scheme is trivially
$\lambda$-multiplicative and strongly $\lambda$-multiplicative, a
proper composition of Shamir's threshold secret sharing schemes results in
a general construction for both $\lambda$-multiplicative LSSS and
strongly $\lambda$-multiplicative LSSS. Let $Q^\lambda$ be a
straightforward extension of $Q^2$ and $Q^3$, that is, an access
structure $AS$ is $Q^\lambda$ if the player set $P$ cannot be covered by
$\lambda$ sets in
$\mathcal{A}=2^P-AS$. The
following corollary is easy to prove.

\begin{corollary}
Let $AS$ be an access structure over $P$. Then there exists a
$\lambda$-multiplicative (respectively, strongly $\lambda$-multiplicative)
LSSS realizing $AS$ if and only if $AS$ is $Q^{\lambda}$ (respectively,
$Q^{\lambda+1}$).
\end{corollary}

Since a $\lambda$-multiplicative LSSS transforms the products of
$\lambda$ entries into a linear combination of the local products
of shares, it can be used to simplify the secure computation of
sequential multiplications. In particular,  when compared to using
only the multiplicative property (which corresponds to the
case when $\lambda=2$), a $\lambda$-multiplicative LSSS can lead
to reduced round complexity by a factor of $\frac{1}{\log\lambda}$
in certain cases.

We also point out that $Q^\lambda$ is not a necessary condition for
secure computation. Instead, the necessary condition is $Q^2$ for the
passive adversary model,   or $Q^3$ for the active adversary model
\cite{Cramer:General}. The condition $Q^\lambda$ is just a necessary condition
for the existence of $\lambda$-multiplicative LSSS which can be
used to simplify computation. In practice, many threshold
adversary structures satisfy the $Q^\lambda$ condition for some appropriate
integer $\lambda$, and the widely used Shamir's threshold secret sharing
scheme is already $\lambda$-multiplicative. By using this
$\lambda$-multiplicativity, we can get more efficient MPC protocols.
However, since the special case $\lambda=3$ shows a close
relationship with strongly multiplicative LSSS, a
fundamental tool in MPC, this paper focuses on 3-multiplicative
LSSS.

\section{Strong Multiplicativity and 3-Multiplicativity} \label{sec:3-strong}

In this section, we show that strong multiplicativity and
3-multiplicativity are closely related. On the one hand,
given a strongly multiplicative LSSS, there is an {\it efficient}
transformation that converts it to a 3-multiplicative LSSS. On the
other hand, we show that any 3-multiplicative LSSS is a
strongly multiplicative LSSS, but the converse is not true. It should be noted that strong
multiplicativity, as defined, has a combinatorial nature.
The definition of 3-multiplicativity is essentially algebraic, which is typically
easier to verify.

\subsection{From Strong Multiplicativity to 3-Multiplicativity} \label{subsec:s->3}

We show a general method to efficiently build a
3-multiplicative LSSS from a strongly multiplicative LSSS, for all
$Q^3$ access structures. As an extension, the proposed method can also be used
to efficiently build a $(\lambda+1)$-multiplicative LSSS from a
strongly $\lambda$-multiplicative LSSS.

\begin{theorem}\label{prop5}
Let $AS$ be a $Q^3$ access structure and
$\mathcal{M}(\mathcal{K},M,\psi,\vec{e}_1)$ be a strongly
multiplicative LSSS realizing $AS$. Suppose that $\mathcal{M}$
has size $d$ and $|\psi^{-1}(P_i)|=d_i$, for $1\leq i\leq n$. Then there
exists a $3$-multiplicative LSSS for $AS$ of size $O(d^2)$.
\end{theorem}

\begin{proof}
We give a constructive proof. Let $M_{\diamond}$ be
the matrix defined in Section \ref{sec:3-m}, and $\psi'$ be the
induced labeling map on the rows of $M_{\diamond}$. Then we have an
LSSS
$\mathcal{M}_{\diamond}(\mathcal{K},M_{\diamond},\psi',\vec{e}_1)$
that realizes an access structure $AS_{\diamond}$. Because
$\mathcal{M}$ is strongly multiplicative, by Proposition \ref{prop2}
we have $\vec{e}_1\in span\{(M_{\diamond})_{\overline{A}}\}$ for all
$A\not\in AS$. Therefore $\overline{A}\in AS_{\diamond}$ and it
follows that $AS^*\subseteq AS_{\diamond}$, where $AS^*$ denotes the
dual access structure of $AS$, defined by
$AS^*=\{A\subseteq P\mid P-A\not\in AS\}$.

The equality (\ref{eq-product}) in the proof of Proposition
\ref{prop2} shows that the diamond product of two share vectors
equals sharing the product of the two secrets by the MSP
$\mathcal{M}_{\diamond}(\mathcal{K},M_{\diamond},\psi',\vec{e}_1)$,
that is,
\begin{equation*}
(M(s_1,\vec{\rho}'_1)^\tau)\diamond(M(s_2,\vec{\rho}'_2)^\tau)=M_{\diamond}(s_1s_2,\vec{\rho})^\tau, ~\mbox{~for some $\vec{\rho}'_1, \vec{\rho}'_2, \vec{\rho} \in {\cal K}^{l-1}$}.
\end{equation*}
Thus, using a method similar to Nikov {\em et al.} \cite{Nikov:multiplicativeLSSS},
 we can get the
product $(s_1s_2)\cdot s_3$ by sharing $s_3$ through the dual MSP of
$\mathcal{M}_{\diamond}$, denoted by $(\mathcal{M}_{\diamond})^*$.
Furthermore, since $(\mathcal{M}_{\diamond})^*$ realizes the dual
access structure $(AS_\diamond)^*$ and $(AS_\diamond)^*\subseteq
(AS^*)^*=AS$, we can build a 3-multiplicative LSSS by the union of
$\mathcal{M}$ and $(\mathcal{M}_{\diamond})^*$, which realizes the
access structure $AS \cup (AS_\diamond)^*=AS$. Now following the same
method of Cramer {\em et al.} and Fehr
\cite{Cramer:General,Fehr:DualMSP}, we prove the required result via the 
construction below.

Compute the column vector $\vec{v}_0$ as a solution to the equation
$(M_{\diamond})^\tau\vec{v}=\vec{e_1}^\tau$ for $\vec{v}$,
and compute $\vec{v}_1,\ldots,\vec{v}_k$ as a basis of the solution
space to $(M_{\diamond})^\tau\vec{v}=\vec{0}^\tau$. Note that
$(M_{\diamond})^\tau\vec{v}=\vec{e_1}^\tau$ is solvable because
$\vec{e}_1\in span\{(M_{\diamond})_{\overline{A}}\}$ for all
$A\not\in AS$, while $(M_{\diamond})^\tau\vec{v}=\vec{0}^\tau$ may
only have the trivial solution $\vec{v}=\vec{0}$ and $k=0$. Let
\begin{equation*}
M'=\left(\begin{array}{cccccc}m_{11}&\cdots&m_{1l}& & &
\\\vdots&\ddots&\vdots& & & \\m_{d1}&\cdots&m_{dl}& & &
\\\vec{v_0}& &
&\vec{v_1}&\cdots&\vec{v_k}\end{array}\right),
\end{equation*}
where
$\left(\begin{array}{ccc}m_{11}&\cdots&m_{1l}\\\vdots&\ddots&\vdots\\m_{d1}&\cdots&m_{dl}\end{array}\right)=M$
and the blanks in $M'$ denote zeros. Define a labeling map
$\psi''$ on the rows of $M'$ which labels the first $d$ rows of
$M'$ according to $\psi$ and the other $\sum_{i=1}^nd_i^2$ rows
according to $\psi'$.

As mentioned above, $\mathcal{M}'(\mathcal{K},M',\psi'',\vec{e}_1)$
obviously realizes the access structure $AS$. We now verify its
3-multiplicativity.

Let $N=(\vec{v}_0, \vec{v}_1,\ldots,\vec{v}_k)$,  a matrix over
$\mathcal{K}$ with $\sum_{i=1}^nd_i^2$ rows and $k+1$ columns. For
$s_i\in\mathcal{K}$ and
$\vec{\rho}_i=(\vec{\rho}'_i,\vec{\rho}''_i)\in
\mathcal{K}^{l-1}\times \mathcal{K}^k$, $1\leq i\leq 3$, denote
$M'(s_i,\vec{\rho}_i)^\tau=(\vec{u}_i,\vec{w}_i)^\tau$, where
$\vec{u}_i^{\;\tau}=M(s_i,\vec{\rho}'_i)^\tau$ and
$\vec{w}_i^{\;\tau}=N(s_i,\vec{\rho}''_i)^\tau$. We have
\begin{equation*}
\vec{u}_1^{\;\tau}\diamond\;
\vec{u}_2^{\;\tau}=(M(s_1,\vec{\rho}'_1)^\tau)\diamond(M(s_2,\vec{\rho}'_2)^\tau)=M_{\diamond}(s_1s_2,\vec{\rho})^\tau,
\end{equation*}
where
$(s_1s_2,\vec{\rho})=(s_1,\vec{\rho}'_1)\otimes(s_2,\vec{\rho}'_2)$.
Then,
\begin{align*}
(\vec{u}_1\diamond\vec{u}_2)\cdot\vec{w}_3^{\;\tau} &=(s_1s_2,\vec{\rho})(M_{\diamond})^{\tau}\cdot N\left(\begin{array}{c}s_3\\ \\ {\vec{\rho}_3''}^{\;\tau}\end{array}\right) \\
&=(s_1s_2,\vec{\rho})
\left(\begin{array}{cccc}1&0&\cdots&0\\0&0&\cdots&0\\\vdots&\vdots&\ddots&\vdots\\0&0&\cdots&0\end{array}\right)\left(\begin{array}{c}s_3\\ \\ {\vec{\rho}_3''}^{\;\tau}\end{array}\right)\\
&=s_1s_2s_3.
\end{align*}

It is easy to see that
$(\vec{u}_1\diamond\vec{u}_2)\cdot\vec{w}_3^{\;\tau}$ is a
linear combination of the entries from
$(\vec{u}_1\diamond\vec{u}_2)\diamond \vec{w}_3$,  and so is a linear
combination of the entries from $M'(s_1,\vec{\rho}_1)^\tau\diamond
M'(s_2,\vec{\rho}_2)^\tau\diamond M'(s_3,\vec{\rho}_3)^\tau$.

Hence $\mathcal{M}'$ is a 3-multiplicative LSSS for $AS$. Obviously,
the size of $\mathcal{M}'$ is $O(d^2)$, since $d+\sum_{i=1}^nd_i^2 <
d^2 + d.$
$\Box$
\end{proof}

If we replace the matrix $M_{\diamond}$ above by the diamond
product of $\lambda$ copies of $M$,  using an identical argument,
the construction from  Theorem \ref{prop5} gives rise to  a
$(\lambda+1)$-multiplicative LSSS from a strongly
$\lambda$-multiplicative LSSS.

\begin{corollary}\label{coro2}
Let $AS$ be a $Q^{\lambda+1}$ access structure and
$\mathcal{M}(\mathcal{K},M,\psi,\vec{e}_1)$ be a strongly
$\lambda$-multiplicative LSSS realizing $AS$. Suppose the size of
$\mathcal{M}$ is $d$ and $|\psi^{-1}(P_i)|=d_i$, for $1\leq i\leq n$.
Then there exists a $(\lambda+1)$-multiplicative LSSS for $AS$ of
size $O(d^{\lambda})$.
\end{corollary}

\subsection{From 3-Multiplicativity to Strong Multiplicativity}

\begin{theorem}\label{th5}
Any $3$-multiplicative LSSS is strongly multiplicative.
\end{theorem}

\begin{proof}
Let $\mathcal{M}(\mathcal{K},M,\psi,\vec{e}_1)$ be a
3-multiplicative LSSS realizing the access structure $AS$ over
$P$. For all $A\in\mathcal{A}=2^P-AS$, by Proposition \ref{prop1},
we can choose a fixed vector $\vec{\rho}''\in\mathcal{K}^{l-1}$
such that $M_{A}(1,\vec{\rho}'')^\tau=\vec{0}^\tau$.
There exists a recombination vector
$\vec{z}\in\mathcal{K}^{\sum_{i=1}^nd_i^3}$ such that for all
$s,s'\in\mathcal{K}$ and
$\vec{\rho},\vec{\rho}'\in\mathcal{K}^{l-1}$, we have
\begin{equation*}
ss'=\vec{z}(M(s,\vec{\rho})^\tau\diamond M(s',\vec{\rho}')^\tau\diamond M(1,\vec{\rho}'')^\tau).\end{equation*}
Since $M_{A}(1,\vec{\rho}'')^\tau=\vec{0}^\tau$, and
$M_{\overline{A}}(1,\vec{\rho}'')^\tau$ is a constant vector for
fixed $\vec{\rho}''$,  the vector
$\vec{z}'\in\mathcal{K}^{\sum_{P_i\not\in A}d_i^2}$ that satisfies
\begin{equation*}
\vec{z}(M(s,\vec{\rho})^\tau\diamond M(s',\vec{\rho}')^\tau\diamond M(1,\vec{\rho}'')^\tau)=\vec{z}'(M_{\overline{A}}(s,\vec{\rho})^\tau\diamond M_{\overline{A}}(s',\vec{\rho}')^\tau)
\end{equation*}
can be easily determined.
Thus $ss'=\vec{z}'(M_{\overline{A}}(s,\vec{\rho})^\tau\diamond
M_{\overline{A}}(s',\vec{\rho}')^\tau)$. Hence,  $\mathcal{M}$ is
strongly multiplicative.
$\Box$
\end{proof}

Although 3-multiplicative LSSS is a subclass of strongly
multiplicative LSSS, one of the advantages of 3-multiplicativity is
that its verification admits a simpler process. For 3-multiplicativity,
we need only to check that $\vec{e}_1\in span\{(M\diamond M\diamond M)\}$,
while strong multiplicativity requires the verification of
 $\vec{e}_1\in span\{(M\diamond M)_{\overline{A}}\}$ for
{\em all} $A\not\in AS$.

Using a similar argument, the following results for
$(\lambda+1)$-multiplicativity can be proved:
\begin{enumerate}[(i)]
\item A
$(\lambda+1)$-multiplicative LSSS is a strongly
$\lambda$-multiplicative LSSS.
\item A $\lambda$-multiplicative LSSS is a $\lambda'$-multiplicative LSSS,
where $1<\lambda'<\lambda$.
\end{enumerate}

\subsection{An Example of  a Strongly Multiplicative LSSS that is Not
3-Multiplicative} \label{subsec:example}

We  give an example of a strongly multiplicative LSSS that is not
3-multiplicative. It follows that  3-multiplicative LSSS are
strictly contained in the class of strongly multiplicative LSSS. The
construction process is as follows. Start with an LSSS that realizes
a $Q^3$ access structure but is not strongly multiplicative. We then
apply the general construction given in Section \ref{subsec:2.3} to
convert it into a strongly multiplicative LSSS. The resulting LSSS
is however not 3-multiplicative.

Let $P=\{P_1,P_2,P_3,P_4,P_5,P_6\}$ be the set of players.
Consider the access structure $AS$ over $P$ defined by
\begin{equation*}
AS_{min}=\{(1,2),(3,4),(5,6),(1,5),(1,6),(2,6),(2,5),(3,6),(4,5)\},
\end{equation*}
where we use subscript to denote the corresponding player. For
example, $(1,2)$ denotes the subset $\{P_1,P_2\}$. It
is easy to verify that the corresponding adversary structure is
\begin{equation*}
\mathcal{A}_{max}=\{(1,3),(1,4),(2,3),(2,4),(3,5),(4,6)\},
\end{equation*}
and that $AS$ is a $Q^3$ access structure.

Let $\mathcal{K}=\mathbb{F}_2$. Define the matrix $M$ over
$\mathbb{F}_2$ with the labeling map $\psi$ such that
\begin{align*}
M_{P_1}=\left(\begin{array}{ccccc}1&0&1&0&0\\0&0&0&1&0\\0&0&0&0&1\end{array}\right),\;
M_{P_2}=\left(\begin{array}{ccccc}0&0&1&0&0\\0&0&0&1&0\\0&0&0&0&1\end{array}\right),\;
M_{P_3}=\left(\begin{array}{ccccc}1&1&0&0&0\\0&0&0&0&1\end{array}\right),
\\
M_{P_4}=\left(\begin{array}{ccccc}0&1&0&0&0\\0&0&0&1&0\end{array}\right),\;
M_{P_5}=\left(\begin{array}{ccccc}1&1&1&0&0\\1&0&0&1&0\end{array}\right),\;
M_{P_6}=\left(\begin{array}{ccccc}0&1&1&0&0\\1&0&0&0&1\end{array}\right).
\end{align*}
It can be verified that the LSSS
$\mathcal{M}(\mathbb{F}_2,M,\psi,\vec{e}_1)$ realizes the access
structure $AS$. Moreover, for all $A\in\mathcal{A}-\{(1,3),(1,4)\}$,
the constricted LSSS $\mathcal{M}_{\overline{A}}$ is multiplicative.
Thus in order to get a strongly multiplicative LSSS, we just need to
expand $\mathcal{M}$ with multiplicativity when constricted to both
 $\{P_2,P_4,P_5,P_6\}$ and $\{P_2,P_3,P_5,P_6\}$.

Firstly,  consider the LSSS $\mathcal{M}$ constricted to
$P'=\{P_2,P_4,P_5,P_6\}$. Obviously, $\mathcal{M}_{P'}$ realizes
the access structure $AS'_{min}=\{(5,6),(2,6),(2,5),(4,5)\}$, which
is $Q^2$ over $P'$. By the method of Cramer {\em et al.} \cite{Cramer:General}, we can
transform $\mathcal{M}_{P'}$ into the multiplicative LSSS
$\mathcal{M}'_{P'}(\mathbb{F}_2,M',\psi',\vec{e}_1)$ defined as
follows:
\begin{align*}
M'_{P_2}=\left(\begin{array}{ccccccccc}0&0&1&0&0& & & & \\0&0&0&1&0& & & & \\0&0&0&0&1& & & & \\ & & & & &0&1&1&1\\ & & & & &1&1&0&0\\ & & & & &0&0&0&1\end{array}\right),~~~~
M'_{P_4}=\left(\begin{array}{ccccccccc}0&1&0&0&0& & & &
\\0&0&0&1&0& & & & \\ & & & & &0&1&1&1\\ & & & &
&1&0&0&0\end{array}\right),\\
M'_{P_5}=\left(\begin{array}{ccccccccc}1&1&1&0&0& & & & \\1&0&0&1&0&
& & & \\1& & & & &0&1&0&1\\0& & & &
&0&1&0&0\end{array}\right),~~~~
M'_{P_6}=\left(\begin{array}{ccccccccc}0&1&1&0&0& & & &
\\1&0&0&0&1& & & & \\1& & & & &0&0&1&0\\0& & & &
&0&0&0&1\end{array}\right),
\end{align*}
where the blanks in the matrices
denote zeros.

For consistency, we define
\begin{align*}
M'_{P_1} &=(M_{P_1}\;O_{3\times4}),\\
M'_{P_3} &=(M_{P_3}\;O_{2\times4}),
\end{align*}
where $O_{m\times n}$ denotes the $m\times n$
matrix of all zeros. It can be verified that
for the subset $P''=\{P_2,P_3,P_5,P_6\}$, the constricted LSSS
$\mathcal{M}'_{P''}$ is indeed multiplicative. Therefore,
$\mathcal{M}'(\mathbb{F}_2,M',\psi',\vec{e}_1)$ is a strongly
multiplicative LSSS realizing the access structure $AS$.
Furthermore,  it can be verified that $\mathcal{M}'$ is not
3-multiplicative (the verification involves checking a
$443\times 729$ matrix using Matlab).

The scheme
$\mathcal{M}(\mathbb{F}_2,M,\psi,\vec{v}_1)$ given above is the
first example of an LSSS which realizes a $Q^3$ access structure
but is not strongly multiplicative.

\section{Constructions for $3$-multiplicative LSSS} \label{sec:constructions}

It is tempting to find efficient constructions for 3-multiplicative
LSSS. In general, it is a hard problem to construct LSSS with
polynomial size for any specified access structure, and it seems to be an even harder problem
to construct polynomial size 3-multiplicative LSSS with general $Q^3$ access structures. We
mention two constructions for 3-multiplicative LSSS. These constructions are
generally inefficient, which can result in schemes with exponential sizes.
The two constructions are:

\begin{enumerate}
\item The Cramer-Damg{\aa}rd-Maurer construction based on Shamir's
threshold secret sharing scheme \cite{Cramer:General}.

\item The construction given in Subsection \ref{subsec:s->3} based
on strongly multiplicative LSSS.
\end{enumerate}

There exist, however, some efficient LSSS with specific access
structures that are multiplicative or 3-multiplicative. For instance,
Shamir's $t$ out of $n$ threshold secret sharing schemes are multiplicative if $n
\geq 2t+1$,  and 3-multiplicative if $n \geq 3t+1$.

On the other hand, secret sharing schemes from error-correcting codes give good
multiplicative properties. It is well known that a secret sharing scheme from a linear
error-correcting code is an LSSS. We know
that such an LSSS is multiplicative provided the underlying code is a
self dual code \cite{Cramer:OnCodes}. The LSSS from a Reed-Solomon code is $\lambda$-multiplicative if
the corresponding access structure is $Q^{\lambda}$. In this
section, we show the multiplicativity of two other classes
of secret sharing schemes from error-correcting codes:
\begin{enumerate}[(i)]
\item schemes from
Reed-Muller codes are $\lambda$-multiplicative LSSS; and 
\item
schemes from algebraic geometric codes are
$\lambda$-multiplicative ramp LSSS.
\end{enumerate}

\subsection{A Construction from Reed-Muller Codes}

Let $\vec{v}_0,\vec{v}_1,\ldots,\vec{v}_{2^m-1}$ be all the points in
the space $\mathbb{F}_2^{\;m}$. The binary Reed-Muller code
$\mathcal{R}(r,m)$ is defined as follows:
\begin{equation*}
\mathcal{R}(r,m)=\{(f(\vec{v}_0),f(\vec{v}_1),\ldots,f(\vec{v}_{2^m-1}))\mid f\in\mathbb{F}_2[x_1,\ldots,x_m],\;\deg f\leq
r\}.
\end{equation*}

Take $f({\vec{v}_0})$ as  the secret, and $f({\vec{v}_i})$ as the
share distributed to player $P_i$, $1\leq i\leq 2^m-1$. Then
$\mathcal{R}(r,m)$ gives rise to an LSSS for the set of players
$\{P_1,\ldots,P_n\}$,  with the secret-domain being
$\mathbb{F}_2$, where
$n=2^m-1$. For any three codewords
\begin{equation*}
\vec{c}_i=(s_i,s_{i1},\ldots,s_{in})=(f_i(\vec{v}_0),f_i(\vec{v}_1),\ldots,f_i(\vec{v}_n))\in\mathcal{R}(r,m),~~1\leq i\leq 3,
\end{equation*}
it is easy to see that
\begin{align*}
\vec{c}_1\diamond\vec{c}_2\diamond\vec{c}_3 &=(s_1s_2s_3,s_{11}s_{21}s_{31},\ldots,s_{1n}s_{2n}s_{3n}) \\
&=(g(\vec{v}_0),g(\vec{v}_1),\ldots,g(\vec{v}_n))\in\mathcal{R}(3r,m),
\end{align*}
where $g=f_1f_2f_3\in\mathbb{F}_2[x_1,\ldots,x_m]$ and $\deg g\leq 3r$.
From basic results on Reed-Muller codes \cite{Lint:Code}, we know that
$\mathcal{R}(3r,m)$ has dual code
$\mathcal{R}(m-3r-1,m)$ when $m>3r$, and  the dual code
$\mathcal{R}(m-3r-1,m)$ trivially contains the codeword
$(1,1,\ldots,1)$. It follows that
$s_1s_2s_3=\sum_{j=1}^ns_{1j}s_{2j}s_{3j}$, which shows that the LSSS from
$\mathcal{R}(r,m)$ is 3-multiplicative when
$m>3r$. Certainly, this LSSS is strongly multiplicative. In general,
we have the following result:

\begin{theorem} \label{thm:RM-code}
The LSSS constructed above from
$\mathcal{R}(r,m)$ is $\lambda$-multiplicative, provided $m >
\lambda r$.
\end{theorem}

\subsection{A Construction from Algebraic Geometric Codes}

Chen and Cramer \cite{CC:AGRamp} constructed secret sharing
schemes from algebraic geometric (AG) codes. These schemes
are {\em quasi-threshold} (or {\em ramp}) schemes, which
means that any $t$ out of $n$ players can recover the secret, and any fewer
than $t'$ players have no information about the secret, where
$t'\leq t \leq n$. In this section, we show that ramp schemes from
some algebraic geometric codes \cite{{CC:AGRamp}} are
$\lambda$-multiplicative.

Let $\chi$ be an absolutely irreducible, projective, and nonsingular
curve defined over $\mathbb{F}_q$ with genus $g$, and let
$D=\{v_0,v_1,\ldots,v_n\}$ be the set of $\mathbb{F}_q$-rational points
on $\chi$. Let $G$ be an $\mathbb{F}_q$-rational divisor with degree $m$
satisfying $supp(G) \cap D=\emptyset$ and $2g-2<m<n+1$. Let
$\overline{\mathbb{F}}_q$ denote the algebraic closure of
$\mathbb{F}_q$, let $\overline{\mathbb{F}}_q(\chi)$ denote the function
field of the curve $\chi$, and let $\Omega(\chi)$ denote all the
differentials on $\chi$. Define the linear spaces:
\begin{align*}
\mathcal{L}(G) &=\{f\in\overline{\mathbb{F}}_q(\chi)\mid (f)+G\geq 0\},\\
\Omega(G) &=\{\omega\in\Omega(\chi)\mid (\omega)\geq G\}.
\end{align*}
Then the functional AG code $C_\mathcal{L}(D,G)$ and residual AG
code $C_{\Omega}(D,G)$ are respectively defined as follows:
\begin{align*}
C_\mathcal{L}(D,G) &=\{(f(v_0),f(v_1),\ldots,f(v_n))\mid f\in\mathcal{L}(G)\}\subseteq \mathbb{F}_q^{\;n+1}, \\
C_{\Omega}(D,G) &=\{(Res_{v_0}(\eta),Res_{v_1}(\eta),\ldots,Res_{v_n}(\eta))\mid\eta\in\Omega(G-D)\}\subseteq\mathbb{F}_q^{\;n+1},
\end{align*}
where $Res_{v_i}(\eta)$ denotes the residue of $\eta$ at $v_i$.

As above, $C_{\Omega}(D,G)$ induces an LSSS for the set of players
$\{P_1,\ldots,P_n\}$, where for every codeword
$(f(v_0),f(v_1),\ldots,f(v_n))\in
C_{\Omega}(D,G)=C_{\mathcal{L}}(D,D-G+(\eta))$, $f(v_0)$ is the
secret and $f(v_i)$ is $P_i$'s share, $1\leq i\leq n$. For any
$\lambda$ codewords
\begin{align*}
\vec{c}_i &=(s_i,s_{i1},\ldots,s_{in}) \\
&=(f_i(v_0),f_i(v_1),\ldots,f_i(v_n))\in C_{\mathcal{L}}(D,D-G+(\eta)),~~~1\leq
i\leq \lambda,
\end{align*}
it is easy to see that
\begin{equation*}
\diamond_{i=1}^\lambda\vec{c}_i=\left(\prod_{i=1}^\lambda s_i,\prod_{i=1}^\lambda s_{i1},\ldots,\prod_{i=1}^\lambda s_{in}\right)\in C_\mathcal{L}(D,\lambda(D-G+(\eta))).
\end{equation*}

If $2g-2<\deg(\lambda(D-G+(\eta)))<n$, then
$C_\mathcal{L}(D,\lambda(D-G+(\eta)))$ has the dual code
$C_{\Omega}(D,\lambda(D-G+(\eta)))=C_{\mathcal{L}}(D,\lambda
G-(\lambda-1)(D+(\eta)))$. When $\deg(\lambda
G-(\lambda-1)(D+(\eta)))\geq 2g$,
$C_{\Omega}(D,\lambda(D-G+(\eta)))$ has a codeword with a nonzero
first coordinate, implying $\prod_{i=1}^\lambda
s_i=\sum_{j=1}^na_j\prod_{i=1}^\lambda s_{ij}$ for some constants
$a_j\in\mathbb{F}_q$. Thus, the LSSS induced by the AG code
$C_{\Omega}(D,G)$ is $\lambda$-multiplicative. It is easy to see that
if $\deg G = m \geq\frac{(\lambda-1)(n-1)}{\lambda}+2g$ then we have
$2g-2<\deg(\lambda(D-G+(\eta)))<n$ and  $\deg(\lambda
G-(\lambda-1)(D+(\eta)))\geq 2g$. Therefore, we have the
following theorem.

\begin{theorem}
Let $\chi$ be an absolutely irreducible, projective, and nonsingular
curve defined over $\mathbb{F}_q$ with genus $g$, let
$D=\{v_0,v_1,\ldots,v_n\}$ be the set of $\mathbb{F}_q$-rational points
on $\chi$. Let $G$ be an $\mathbb{F}_q$-rational divisor with degree $m$
satisfying $supp(G) \cap D=\emptyset$ and $2g-2<m<n+1$. Then the
LSSS induced by the AG code $C_{\Omega}(D,G)$ is
$\lambda$-multiplicative, provided
$m\geq\frac{(\lambda-1)(n-1)}{\lambda}+2g$.
\end{theorem}

\section{Implications of the Multiplicativity of LSSS} \label{sec:implication}

The property of 3-multiplicativity implies strong multiplicativity, and so is
sufficient for building MPC protocols against active adversaries. 
The conditions for 3-multiplicativity are
easy to verify, while verification for strong multiplicativity
involves checking an exponential number of  equations (each subset in
the adversary structure corresponds to an equation).

With 3-multiplicative LSSS, or more generally
$\lambda$-multiplicative LSSS, we can simplify local computation
for each player and reduce the round complexity in MPC
protocols. For example, using the technique of Bar-Ilan and Beaver
\cite{BB:ConstantRd}, we can compute $\prod_{i=1}^lx_i,\;
x_i\in\mathbb{F}_q$, in a constant number of rounds, independent of $l$. For
simplicity, we consider passive adversaries in the
information-theoretic model. Suppose for $1\leq i\leq l$, the shares
of $x_i$, denoted by $[x_i]$, have already been distributed among
the players. To compute $\prod_{i=1}^lx_i,\; x_i\in\mathbb{F}_q$, we
follow the process of Cramer {\em et al.} \cite{CKP:LinearAlgbra}:
\begin{enumerate}[(1)]
\item Generate
$[b_0\in_R\mathbb{F}_q^{\;*}],[b_1\in_R\mathbb{F}_q^{\;*}],\ldots,[b_l\in_R\mathbb{F}_q^{\;*}]$
and $[b_0^{-1}],[b_1^{-1}],\ldots,[b_l^{-1}]$, where
$b_i\in_R\mathbb{F}_q^{\;*}$ means that $b_i$ is a random element in
$\mathbb{F}_q^{\;*}$.

\item For $1\leq i\leq l$, each player computes $[b_{i-1}x_ib_i^{-1}]$
from $[b_{i-1}],[b_i^{-1}]$ and $[x_i]$.

\item Recover $d_i=b_{i-1}x_ib_i^{-1}$ from $[b_{i-1}x_ib_i^{-1}]$ for
$1\leq i\leq l$, and compute $d=\prod_{i=1}^{l}d_i$.

\item Compute $[db_0^{-1}b_l]$ from $[b_0^{-1}],[b_l]$ and $d$.
\end{enumerate}

It is easy to see that $db_0^{-1}b_l=\prod_{i=1}^lx_i$. Using
a strongly multiplicative LSSS, the above process takes five
rounds of interactions as two rounds are required in Step (2).
However,  if we use a 3-multiplicative LSSS instead, then
only one round is needed for Step (2). Thus, 3-multiplicative LSSS reduce
the round complexity of computing unbounded fan-in multiplication
from five to four. This in turn simplifies the computation of many
problems, such as polynomial evaluation and solving linear
systems of equations.

In general, the relationship between  $\lambda$-multiplicative LSSS
and strongly $\lambda$-multiplicative LSSS can be described as
follows:
\begin{equation*}
\cdots\subseteq SMLSSS_{\lambda+1}\subsetneq  MLSSS_{\lambda+1}\subseteq SMLSSS_{\lambda}\subsetneq MLSSS_{\lambda}\subseteq\cdots,
\end{equation*}
where $MLSSS_{\lambda}$ (respectively, $SMLSSS_{\lambda}$) denotes the class
of  $\lambda$-multiplicative (respectively, strongly $\lambda$-multiplicative) 
LSSS. It is easy to see that $SMLSSS_{\lambda}\subsetneq
MLSSS_{\lambda}$ because they exist under the conditions
$Q^{\lambda+1}$ and $Q^\lambda$, respectively. Since
$SMLSSS_{\lambda}$ and $MLSSS_{\lambda+1}$ both exist under the same
necessary and sufficient condition of $Q^{\lambda+1}$, it is not
straightforward to see whether $MLSSS_{\lambda+1}$ is strictly
contained in $SMLSSS_{\lambda}$. For $\lambda=2$, we already know that
$MLSSS_3\subsetneq SMLSSS_2$ (Section \ref{subsec:example}). It
would be interesting to find out if this is also true for $\lambda >
2$. We have also given an efficient transformation from
$SMLSSS_{\lambda}$ to $MLSSS_{\lambda+1}$. It remains open whether an
efficient transformation from $MLSSS_{\lambda}$ to
$SMLSSS_{\lambda}$ exists when
the access structure is $Q^{\lambda+1}$. When $\lambda=2$, this is
a well-known open problem \cite{Cramer:General}.

\section{Conclusions} \label{sec:conclusion}

In this paper, we propose the new concept of 3-multiplicative LSSS, which form a
subclass of strongly multiplicative LSSS. The 3-multiplicative LSSS are easier to construct
compared to strongly multiplicative LSSS. 
They
can also simplify the computation and reduce the round complexity in
secure multiparty computation protocols. We believe that
3-multiplicative LSSS are a more appropriate primitive as
building blocks for secure multiparty computations, and deserve
further investigation. We stress that finding efficient
constructions of 3-multiplicative LSSS for general access
structures remains an important open problem.

\section*{Acknowledgement}
The work of M. Liu and Z. Zhang is supported in part by the 973 project of China
(No. 2004CB318000). Part of the work was done while Z. Zhang was visiting Nanyang Technological University
supported by the Singapore Ministry of Education under Research Grant T206B2204.

The work of Y. M. Chee, S. Ling, and H. Wang is supported in part by
the Singapore National Research Foundation under Research Grant NRF-CRP2-2007-03.

In addition, the work of Y. M. Chee
is also supported in part by the Nanyang
Technological University under Research Grant M58110040, and the work of
H. Wang is also supported in part by
the Australian Research Council under ARC Discovery Project
DP0665035.


\begin{thebibliography}{1000}
\bibitem{BB:ConstantRd}
J. Bar-Ilan , D. Beaver, Non-cryptographic fault-tolerant computing
in constant number of rounds of interaction. PODC'89,
pp. 201-209, 1989.

\bibitem{Beimel:Secure}
A. Beimel, Secure schemes for secret sharing and key distribution.
PhD thesis, Technion - Israel Institute of Technology, 1996.

\bibitem{CC:AGRamp}
H. Chen, R. Cramer, Algebraic geometric secret sharing schemes and
secure multi-party computations over small fields. CRYPTO'06, LNCS, 
vol. 4117, pp. 521-536, 2006.

\bibitem{CKP:LinearAlgbra}
R. Cramer, E. Kiltz, C. Padr\'{o}, A note on secure computation of
the Moore-Penrose pseudoinverse and its spplication to secure linear
algebra. CRYPTO'07, LNCS, vol. 4622, pp. 613-630, 2007.

\bibitem{CC:SMRamp}
H. Chen, R. Cramer, R. de Haan and I. Cascudo Pueyo. Strongly
multiplicative ramp schemes from high degree rational points on
curves.  EUROCRYPT'08, LNCS, vol. 4965, pp. 451-470, 2008.

\bibitem{Cramer:General}
R. Cramer, I. Damg{\aa}rd,  U. Maurer,  General secure multi-party
computation from any linear secret-sharing scheme. EUROCRYPT'00, LNCS, vol. 1807, pp. 316-334, 2000.

\bibitem{Cramer:OnCodes}
R. Cramer, V. Daza, I. Gracia, J. Urroz, G. Leander, J.
Mart\'{\i}-Farr\'{e}, C. Padr\'{o}, On codes, matroids and secure
multi-party computation from linear secret sharing schemes. CRYPTO'05,
LNCS, vol. 3621, pp. 327-343, 2005.

\bibitem{Fehr:DualMSP}
S. Fehr, Efficient construction of the dual span program. Master Thesis, the Swiss Federal Institute of Technology (ETH) Z\"{u}rich, 1999\\
http://homepages.cwi.nl/$\sim$fehr/publications.html 

\bibitem{Goldreich:How}
O. Goldreich, S. Micali, A. Wigderson, How to play ANY mental game. STOC'87, pp. 218-219, 1987.


\bibitem{Karchmer:On}
M. Karchmer, A. Wigderson, On span programs. Proc. 8th Ann. Symp.
Structure in Complexity Theory, pp. 102-111, 1993.

\bibitem{KNN:SMHTSS}
E. K\"{a}sper, V. Nikov, S. Nikova, Strongly multiplicative
hierarchical threshold secret sharing. In 2nd International
Conference on Information Theoretic Security - ICITS 2007, LNCS,
to appear.

\bibitem{Liu:IEEE}
M. Liu, L, Xiao, Z. Zhang, Multiplicative linear secret sharing
schemes based on connectivity of graphs. IEEE Transactions on
Information Theory 53(11), pp. 3973-3978, 2007.

\bibitem{Massey:codeSS93}
J.L. Massey, Minimal codewords and secret sharing. Proc. 6th Joint
Swedish-Russian Workshop on Information Theory, pp. 276-279, 1993.

\bibitem{Nikov:multiplicativeLSSS}
V. Nikov, S. Nikova, B. Preneel, On multiplicative linear secret
sharing schemes. Indocrypt'03, LNCS, vol. 2904, pp. 135-147, 2003.

\bibitem{Lint:Code}
J.H. van Lint, Introduction to coding theory. 3rd edition, Graduate
Texts in Mathematics 86, Springer, 1999. 

\bibitem{Yao:Protocols}
A. Yao, Protocols for secure computation. FOCS '82,
pp. 160-164, 1982.


\end{thebibliography}
\end{document}